\documentclass[aps,twocolumn,showpacs]{revtex4}

\usepackage{mathptm}    
\usepackage{dcolumn}                    
\usepackage{bm}                         
\usepackage{bbm}
\usepackage{dsfont}
\usepackage{graphicx}
\usepackage{amsmath}
\usepackage{feynmp}

\usepackage{appendix}
\usepackage{subfigure}

\newcommand {\ba} {\begin{eqnarray}}
\newcommand {\ea} {\end{eqnarray}}

\usepackage{times}

\begin{document}

\title{Phonon Boost Effect on the $S^{\pm}$-wave Superconductor with Incipient Band}

\author{Yunkyu Bang$^{1,2}$}
\email[]{ykbang@apctp.org}
\affiliation{$^1$Asia Pacific Center for Theoretical Physics, Pohang, Gyeongbuk 790-784, Korea \\
$^2$Department of Physics, POSTECH, Pohang, Gyeongbuk 790-784, Korea}

\begin{abstract}
We showed that the all phonons -- not only forward-scattering phonon but also local (all-momentum-scattering) phonon -- contribute to boosting $T_c$ of the $s^{\pm}$-wave pairing state in the incipient band model. In particular, when the incipient band sinks deeper, the phonon boost effect of the local phonon increases and becomes as effective as the one of the forward-scattering phonon.
Our finding implies that all interface phonons -- not only the 90 $meV$ Fuchs-Kliewer (F-K) phonon but also the 60 $meV$ F-K phonon -- from the SrTiO$_3$ substrate and all intrinsic phonons of the FeSe monolayer itself should contribute to increase $T_c$ of the FeSe/STO monolayer system.
\end{abstract}


\date{\today}
\maketitle

\section{Introduction.}
The discovery of the FeSe/SrTiO$_3$ monolayer system ($T_c \approx 60-100K$) \cite{FeSe1,FeSe2,FeSe3} is posing a serious challenge to our understanding of the Iron-based superconductors (IBS)\cite{Greg_RMP,Hirschfeld,Hirschfeld2}. As to the origin of such a high $T_c$ value,
it is widely accepted that the forward-scattering phonon, penetrating from the SrTiO$_3$ substrate, is the key booster for increasing $T_c$\cite{FeSe_phonon,DHLee_phonon,DHLee_phonon2,Johnston,Johnston2}.  Indeed, Lee {\it et al.}\cite{FeSe_phonon} have measured the replica band shifted by about $90 meV$ downward from the original electron band, and it is claimed to be a strong evidence for the presence of the forward-scattering phonon coupled to the conduction band electrons in the FeSe monolayer. Incidentally, this phonon energy coincides with one of the Fuchs-Kliewer (F-K) modes measured at the interface of the SrTiO$_3$(STO) substrate\cite{FK_phonon_2016}.

However, recently Sawatzky and coworkers\cite{sawatzky2017} have argued that the replica band observed in the Angle Resolved Photo-Emission Spectroscopy (ARPES) experiment\cite{FeSe_phonon} is not the evidence of a forward-scattering phonon coupled to the conduction electrons of the FeSe layer but a consequence of the kinematics of the escaping electrons in the ARPES measurement. The main point of this criticism is that while it is true that there exists a F-K phonon mode of the energy $\sim 90 meV$, there is no evidence that this phonon couples to the conduction band electrons of the FeSe layer dominantly with the small momentum exchange (forward-scattering).

In this paper, we studied the effect of the generic types of phonons on the superconducting (SC) instability of the $s^{\pm}$-gap symmetry in the incipient band model. We found that the all phonons -- not only forward-scattering phonon but also local (all-momentum-scattering) phonon -- contribute to boost $T_c$ of the incipient band superconductor. In particular, we showed that when the incipient band sinks deeper, the phonon boost effect of both types of phonon becomes indistinguishably similar from one another. Therefore, the sunken hole band plays an active role to turn otherwise useless phonons (all-momentum scattering phonons) into useful pairing glues.
Our finding implies that all phonons, both interface phonons -- not only the 90 $meV$ F-K phonon but also the 60 $meV$ F-K phonon -- from the STO substrate and all intrinsic phonons of the FeSe monolayer itself should contribute to increase $T_c$ of the FeSe/STO monolayer system.

\section{Phonon Boost Mechanism}
The basic reason for this surprising result is because the relative size of $\Delta^{+}$-gap and $\Delta^{-}$-gap of the $s^{\pm}$-wave pairing state in the incipient band superconductor is not equal, and this size disparity grows as the incipient band sinks deeper.
To illustrate the consequence of this effect, let us recollect the general principle of the optical phonon contribution to the unconventional superconductor with a sign-changing order parameter (OP).

In the BCS pairing theory, the gap equation at $T_c$ with a gap function $\Delta(k)$ has the following structure

\begin{eqnarray}
\Delta(k) &=& - \sum_{k'} V_{sf}(k,k') \Delta(k') \chi_{sf}(T) \nonumber \\
        & & - \sum_{k'} V_{ph}(k,k') \Delta(k') \chi_{ph}(T),
\end{eqnarray}
where $V_{sf}(k,k') (>0)$ is a repulsive spin-fluctuation mediated interaction and $V_{ph}(k,k') (<0)$ is an attractive phonon interaction.
$\chi_{sf(ph)}(T)$ are the pair susceptibilities defined as $\chi_{sf}(T)= N(0) \int_{-\Lambda_{sf}}^{\Lambda_{sf}} d \xi \frac{\tanh(\frac{\xi}{2 T})} {2 \xi}$ and $\chi_{ph}(T)= N(0) \int_{-\omega_D}^{\omega_D} d \xi \frac{\tanh(\frac{\xi}{2 T})} {2 \xi} \sim N(0) \phi(T)$, respectively.
Assuming that the gap symmetry is already determined by the primary pairing interaction $V_{sf}(k,k')$, the following quantity defines the additional contribution from the phonon interaction $V_{ph}(k,k')$ to the total pairing,

\begin{equation}
\phi(T) N(0) \sum_{k'} V_{ph}(k,k') \Delta(k') = \phi(T) N(0) < V_{ph}(k,k') \Delta(k')>_{k'},
\end{equation}
where $<\cdots >_k$ means the Fermi surface (FS) average.
For example, the contribution of the local Einstein phonon interaction $V_{ph}(k,k')=V_0$ to the $d$-wave gap, $\ \Delta_d(k) \sim (\cos k_x -\cos k_y )$, would be null because
\begin{equation}
V_0 <\Delta_d(k')>_{k'} =0.
\end{equation}
On the other hand, if the phonon potential $V_{ph}(k,k')$ represents a forward-scattering phonon, namely, a dominantly stronger potential when the angle between $\vec{k}$ and $\vec{k'}$ is smaller than a certain angle, say, $\theta < \pi/4$, it is obvious that
\begin{equation}
<V_{ph}(k,k') \Delta_d(k')>_{k'} \neq 0,
\end{equation}
so that the attractive phonon interaction cooperates with the repulsive spin fluctuation interaction $V_{sf}(k,k')$ in Eq.(1) to boost $T_c$ of the $d$-wave pairing $\Delta_d(k)$ \cite{Bang_phonon1}.

The same mechanism would work for the $s^{\pm}$-wave pairing state. Here the relevant quantity is
\begin{equation}
\sum_{k'} V_{ph}(k,k') [\Delta^{+}(k') + \Delta^{-}(k')],
\end{equation}
where $\Delta^{+}(k)$ and $\Delta^{-}(k)$ are the gap functions on each of the hole and electron bands.
Assuming the same size of OPs $|\Delta^{+}(k')|=|\Delta^{-}(k')|$ but with opposite signs, then the contribution of the local Einstein phonon interaction $V_{ph}(k,k')=V_0$ to the $s^{\pm}$-wave pairing is proportional to
\begin{equation}
V_{0} <[\Delta^{+}(k') + \Delta^{-}(k')]>_{k'} =0,
\end{equation}
hence the phonon boost effect of the ordinary Einstein phonon is null for the standard $s^{\pm}$-wave state.
On the other hand, with a forward-scattering phonon, namely, a dominantly stronger potential when $\Delta k =|\vec{k}-\vec{k'}|$ is smaller than the typical distance between the hole band and electron band in the Brillouin zone (BZ), say, $\Delta k < Q =|(\pi,\pi)|$, it is obvious that
\begin{equation}
<V_{ph}(k,k') [\Delta^{+}(k'_h) + \Delta^{-}(k'_e)]>_{k'} \neq 0,
\end{equation}
for any fixed momentum $k$ either on the hole FS or on the electron FS. As a result, the forward-scattering phonon interaction boosts the $T_c$ of the $s^{\pm}$-wave pairing state\cite{Bang_phonon2} as in the case of the $d$-wave state.

Having illustrated the above cases, it is easy to write down the most general condition for gaining the phonon boost effect for the $s^{\pm}$-wave pairing state as follows,
\begin{equation}
<V_{ph}(k,k') [\Delta^{+}(k'_h)\chi_{ph}^h(T)  + \Delta^{-}(k'_e) \chi_{ph}^e (T) ]>_{k'} \neq 0.
\end{equation}
It is well known that the sizes of the gap $|\Delta^{+}|$ and $|\Delta^{-}|$ are not equal in general when $N_{h}(0)\neq N_{e}(0)$ \cite{Bang_pairing}. Furthermore, in the case of the incipient band model\cite{Bang_shadow,Hirschfeld_incipient,Bang_RG}, the pair susceptibility $\chi_{ph}^{h(e)}$ for the hole and electron band can be very different because the integration ranges are different such as $\chi_{ph}^h\sim \int_{-\omega_D}^{-\epsilon_b} d \xi \cdots$ and $\chi_{ph}^e \sim \int_{-\omega_D}^{\omega_D} d \xi \cdots$, respectively, where $\epsilon_b$ is the incipient band distance (see Fig.1).
Therefore, in the incipient band superconductor, Eq.(8) can be largely deviated from zero regardless of whether the phonon potential $V_{ph}(k,k')$ is a forward-scattering phonon or a local (all-momentum-scattering) phonon. As a result, we expect that all types of phonons would contribute to increasing the $T_c$ of the $s^{\pm}$-wave state in the incipient band model\cite{exception}.
In the following, we studied this effect quantitatively with numerical calculations of $T_c$ of a minimal incipient band model, and confirmed that our expectation is indeed true.

\begin{figure}
\noindent
\includegraphics[width=95mm]{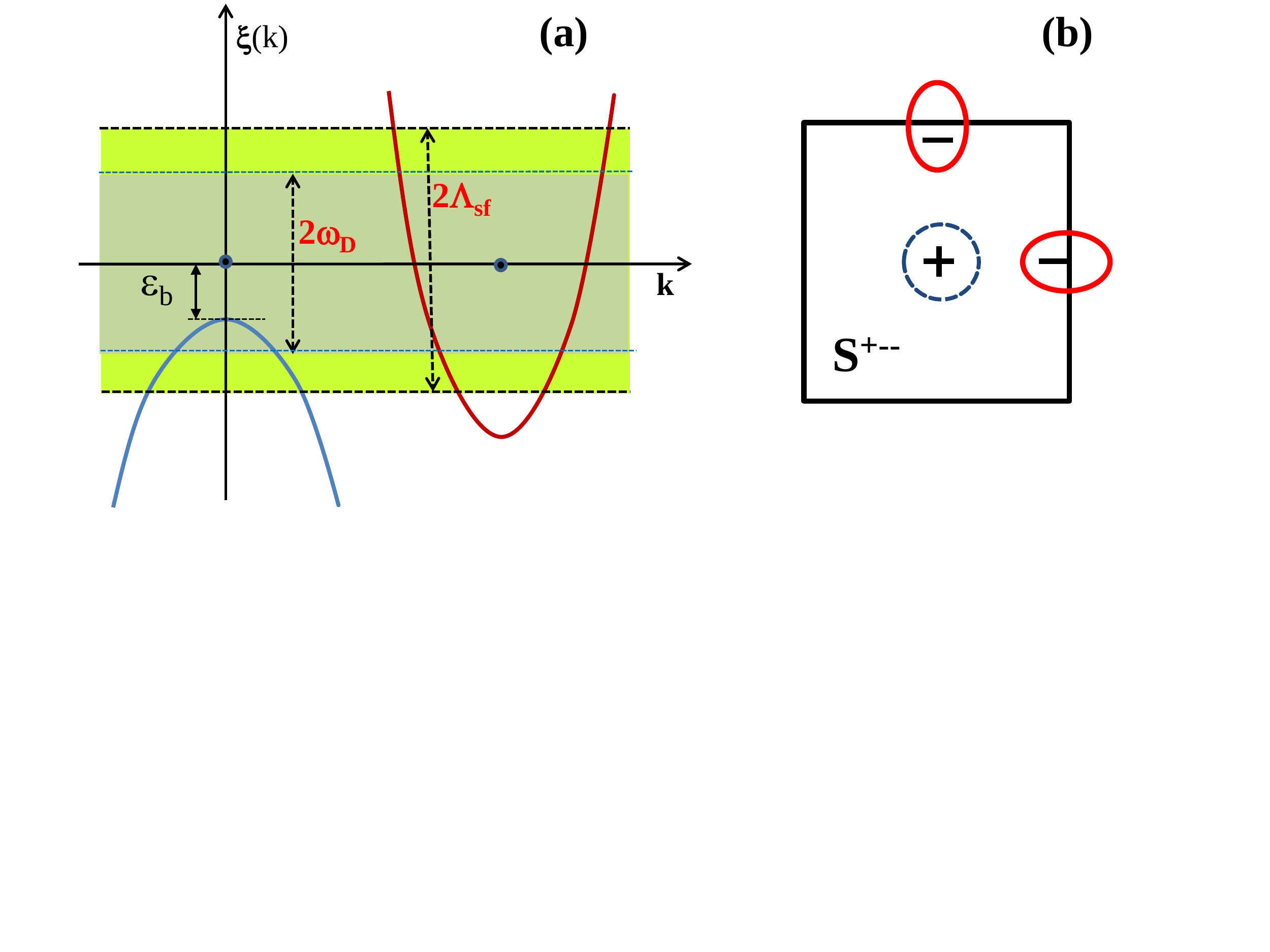}
\vspace{-3.5cm}
\caption{(Color online) (a) A typical incipient band model with $\omega_D < \Lambda_{sf}$. The phonon energy cutoff $\omega_D$ can be larger or smaller than $\epsilon_b$. (b) Schematic picture of the Fermi surfaces and the incipient $s_{he}^{+-}$-wave solution. The hole band has no FS and the dotted circle at $\Gamma$ point only indicates the SC gap character.
\label{fig1}}
\end{figure}

\section{Incipient Band Model}

The minimal incipient band model is depicted in Fig.1. The hole band is sunken below the Fermi level by $\epsilon_b$, hence it has no Fermi surface (FS), and the two electron bands located at X and Y points are treated as one electron band.

For the pairing interactions, we assumed that the spin fluctuation mediated repulsive interaction $V_{sf}(k,k') (>0)$ is operating within the cutoff energy scale $\Lambda_{sf}$ and the phonon mediated attractive interaction $V_{ph}(k,k') (<0)$ is operating within the cutoff energy scale $\omega_{D}$ ($< \Lambda_{sf}$). This model has the incipient $s^{\pm}$-wave solution as the best SC ground state \cite{Bang_shadow, Hirschfeld_incipient}as depicted in Fig.1(b).

For simplicity of calculations but without loss of generality, we simplify the momentum dependent pairing interactions $V_{sf(ph)}(k,k')$ as the $2\times2$ matrix potentials depicting the intra-band and inter-band interactions $V_{sf(ph)}^{ab}, (a,b = h,e)$, then the $T_c$-equation of the incipient two band model is written as

\begin{eqnarray}
\Delta_h  &=&     \bigl[ V^{hh}_{sf} \chi^h_{sf} + V^{hh}_{ph} \chi^h_{ph}  \bigr] \Delta_h +
    \bigl[ V^{he}_{sf}  \chi^e_{sf} + V^{he}_{ph}  \chi^e_{ph}  \bigr] \Delta_e , \\ \nonumber
\Delta_e  &=&     \bigl[ V^{ee}_{sf} \chi^e_{sf} + V^{ee}_{ph} \chi^e_{ph}  \bigr] \Delta_e +
\bigl[ V^{eh}_{sf}  \chi^h_{sf} + V^{eh}_{ph}  \chi^h_{ph}  \bigr] \Delta_h
\end{eqnarray}
where the pair susceptibilities are defined as
\begin{eqnarray}
\chi^{h}_{sf(ph)}(T) &=&  - \frac{N_h}{2} \int _{-\Lambda_{sf(ph)}} ^{-\epsilon_{b}} \frac{d\xi}{\xi}
\tanh (\frac{\xi}{2 T}) \\ \nonumber
\chi^{e}_{sf(ph)}(T) &=&  - N_e \int _{-\Lambda_{sf(ph)}} ^{\Lambda_{sf(ph)}} \frac{d\xi}{\xi}
\tanh (\frac{\xi}{2 T}) \\ \nonumber
\end{eqnarray}
where $\Lambda_{ph}=\omega_D$. Obviously $\chi^{h}_{ph}(T)=0$ when $\omega_D < \epsilon_b$.

\vspace{0cm}
\begin{figure}
\noindent
\includegraphics[width=130mm]{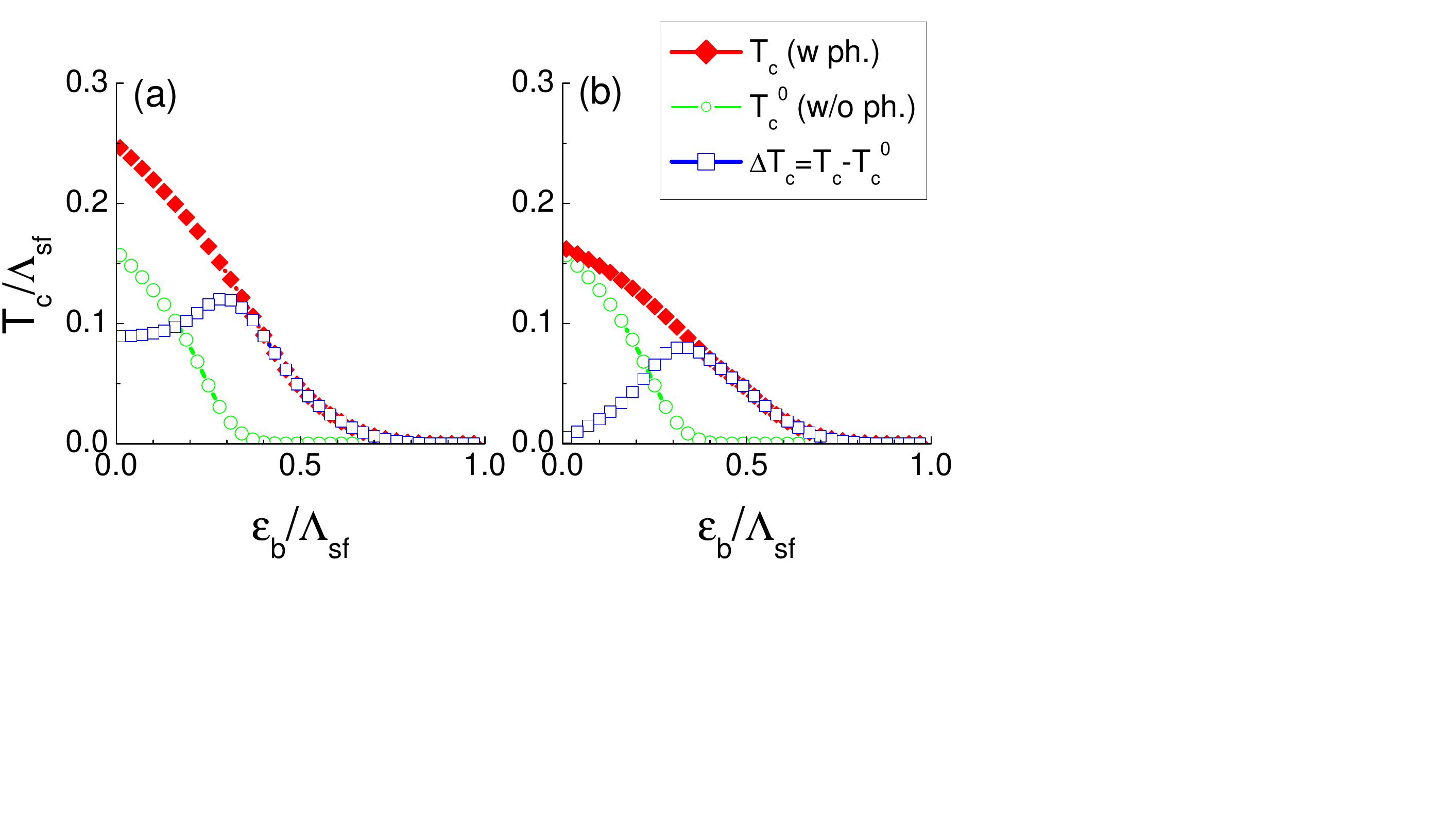}
\vspace{-2.5cm}
\caption{(Color online) (a) Calculated $T_c$ {\it vs} $\epsilon_b$ of Eq.(9). Green symbols are $T_c^0$ without the phonon interaction, i.e. $V_{ph}^{ab}=0$, but only with the spin fluctuation mediated repulsive potential $\sqrt{N_e N_h} V^{he (eh)}_{sf}=1.5$ and $N_{e(h)} V^{ee (hh)}_{sf}=0.4$. Red diamond symbols are $T_c$ including the additional forward-scattering phonon interaction $N_e V^{ee (hh)}_{ph} = -0.5$ and $\sqrt{N_h N_e} V^{he (eh)}_{ph}=0.0$. The blue square symbols are the net phonon boost effect $\Delta T_c = T_c -T_c^0$.
(b) The same calculations as (a) but with the all-momentum-scattering local phonon interaction $N_e V^{ee (hh)}_{ph} = \sqrt{N_h N_e} V^{he (eh)}_{ph} = -0.5$. \label{fig2}}
\end{figure}

\section{Results and Discussions}
First, we calculated the $T_c^0$ as a function of $\epsilon_b$ without phonon interaction, i.e. $V_{ph}^{ab}=0$, but only with spin fluctuation mediated repulsive potential $V^{ab}_{sf} (>0)$. We use representative values of the spin-fluctuation mediated repulsive potential $\sqrt{N_e N_h} V^{he (eh)}_{sf}=1.5$ and $N_{e(h)} V^{ee (hh)}_{sf}=0.4$, and assumed $N_e=N_h$ in all our calculations in this paper. The result of $T_c^0$ is the green circle symbols in Fig.2(a) and (b). $T_c^0$ gradually decreases as $\epsilon_b$ increases as expected \cite{Bang_shadow,Hirschfeld_incipient}.

Now we turn on the attractive phonon interaction $V^{ab}_{ph} (<0)$ in addition to the repulsive spin fluctuation interaction $V_{sf}^{ab}$. We assume the Debye frequency $\omega_D = 0.5\Lambda_{sf}$ for all calculations in this paper.
We first test a forward-scattering phonon, i.e. $N_e V^{he (eh)}_{ph}=0.0$ and $N_e V^{ee (hh)}_{ph} = -0.5$. The results of the calculated $T_c$ are the solid diamond symbols (red) in Fig.2(a). Apparently, $T_c$ is enhanced almost uniformly from $T_c^0$. To see more details, we extracted the net amount of the phonon boost effect of $T_c$ as $\Delta T_c = T_c -T_c^0$, which is plotted by the blue square symbols in Fig.2(a). Interestingly, the phonon boost effect of the purely forward-scattering phonon shows an interesting dependence on $\epsilon_b$; it peaks roughly when the $T_c^0$ collapses to zero.
This behavior tells us a complicated role of the phonon interaction for the total pairing instability. First, the fact that $\Delta T_c$ is always $>0$ definitely proves that the phonon interaction cooperates with the spin-fluctuation mediated interaction to increase $T_c$. However, the fact that $\Delta T_c$ has a maximum peak near when $T_c^0$ approaches zero indicates that there is also a partial competition (or cancellation) between the attractive phonon interaction $V_{ph}^{ab} (<0)$and the repulsive spin-fluctuation mediated interaction $V_{sf}^{ab} (>0)$. However, this partial competition is very weak for the forward-scattering phonon.
Besides this detail, the results of Fig.2(a) confirms the well known concept of the forward-scattering phonon boost effect of $T_c$ in the unconventional superconductor with a sign-changing gap function.

Next, we test an Einstein local phonon (all-momentum-scattering phonon), i.e., $\sqrt{N_e N_h} V^{he (eh)}_{ph}=N_{e(h)} V^{ee (hh)}_{ph} = -0.5$. The calculated $T_c$ is plotted in Fig.2(b) as the red diamond symbols. It shows that the phonon boost effect $\Delta T_c = T_c -T_c^0$ (blue squares) is about the same magnitude as the pure forward-scattering phonon except the region where $\epsilon_b$ is small. When $\epsilon_b$ is small, the sizes of the OPs are close each other as $|\Delta_h^{+}| \sim |\Delta_e^{-}|$, so that the phonon contribution of Eq.(8) becomes close to 0, hence the phonon-boost effect is very weak. As $\epsilon_b$ increases, the net phonon boost effect $\Delta T_c = T_c -T_c^0$ increases until it reaches the maximum and eventually decreases.
$\Delta T_c$ increases because the gap size disparity $|\Delta_h^{+}| / |\Delta_e^{-}|$ increases as the hole band sinks deeper. Beyond crossing a certain depth as $\epsilon_b > \epsilon_b^{\ast}$, both $\Delta T_c$  and $T_c$ itself decreases because the absolute phase space for the pairing interaction shrinks to zero except the intra-electron-band scattering.
This is a totally unexpected result from the common belief that the {\it "forward-scattering"} is the necessary condition for the phonon boost effect. Our result of Fig.2(b) is a clear demonstration that all phonons should contribute to enhance $T_c$, and the sunken band in the incipient band superconductor plays an active role for this unusual behavior.

To see more details, we calculate the $T_c$ with the different values of the phonon interaction strength for both cases, respectively. The strength of the spin-fluctuation mediated interaction $V_{sf}^{ab}$ is fixed as the same values used in Fig.1 in all calculations.

Fig.3 is the results of $T_c$ with the forward-scattering phonon. The main panel is the calculated $T_c$ with different values of the forward-scattering phonon interaction as $N_e V^{ee (hh)}_{ph} =0.0, -0.5, -1.0, -1.5$, respectively, in increasing order of $T_c$, and $\sqrt{N_h N_e} V^{he (eh)}_{ph}=0.0$ for all cases.
The inset of Fig.3 is the net phonon-boost effect $\Delta T_c = T_c -T_c^0$ for each cases. The overall behavior is similar in all cases. The magnitude of $\Delta T_c$ monotonically increases with the strength of the phonon interaction, and its peak position is always near $\epsilon_b^{\ast}$ where $T_c^0$ approaches zero. One particular thing to note is that there is no noticeable change in $\Delta T_c$ when $\epsilon_b$ crosses the phonon interaction cutoff $\omega_D$, which is not the case for the all-momentum-scattering phonon.

Fig.4 is the results of $T_c$ with the all-momentum-scattering phonon, hence $N_e V^{ee (hh)}_{ph}=\sqrt{N_h N_e} V^{he (eh)}_{ph}=\lambda_{ph}$ for all calculations. The coupling strength increases as $\lambda_{ph} =0.0, -0.5, -1.0, -1.5$ and $-2.0$, respectively, in increasing order of $T_c$. The increased $T_c$ is the similar magnitude as in the case of the froward-scattering phonon except for the small $\epsilon_b$ region. A new finding is that this small $\epsilon_b$ region is defined as $\epsilon_b < \epsilon_b^{\ast}$ for weak coupling phonon (see the $\lambda_{ph}=-0.5$ data in the inset of Fig.4). Increasing the coupling strength, this small $\epsilon_b$ region, where $\Delta T_c$ is increasing as $\epsilon_b$ increases, extends to $\epsilon_b < \omega_D$. This behavior can be clearly seen by comparing the insets of Fig.3 and Fig.4. The $\Delta T_c$ of the forward-scattering phonon case in Fig.3 has always the maximum peak at $\epsilon_b = \epsilon_b^{\ast} \approx 0.32$. On the other hand, the peak position of $\Delta T_c$ of the all-momentum-scattering phonon case in Fig.4 shifts from $\epsilon_b = \epsilon_b^{\ast} \approx 0.32$ to $\epsilon_b = \omega_D=0.5$ as the phonon coupling increases.
As a result, when the phonon coupling strength is strong enough such as $\lambda_{ph}=-1.5$ and $-2.0$, the increasing slope of $\Delta T_c$ is so steep that the total $T_c$ itself develops a maximum peak at $\epsilon_b = \omega_D$ and decreases afterwards. This behavior of $T_c$ {\it vs} $\epsilon_b$ is very different from a standard incipient band model where $T_c$ monotonically decreases as $\epsilon_b$ increases.

One final important remark is that while the calculated $T_c$ in this paper is always of the $s_{he}^{+-}$-gap solution as depicted in Fig.1(b), we have also checked the possibility of the $s_{he}^{++}$-gap solution and it never be a solution for all cases of this paper.\\
\vspace{0cm}
\begin{figure}
\noindent
\includegraphics[width=120mm]{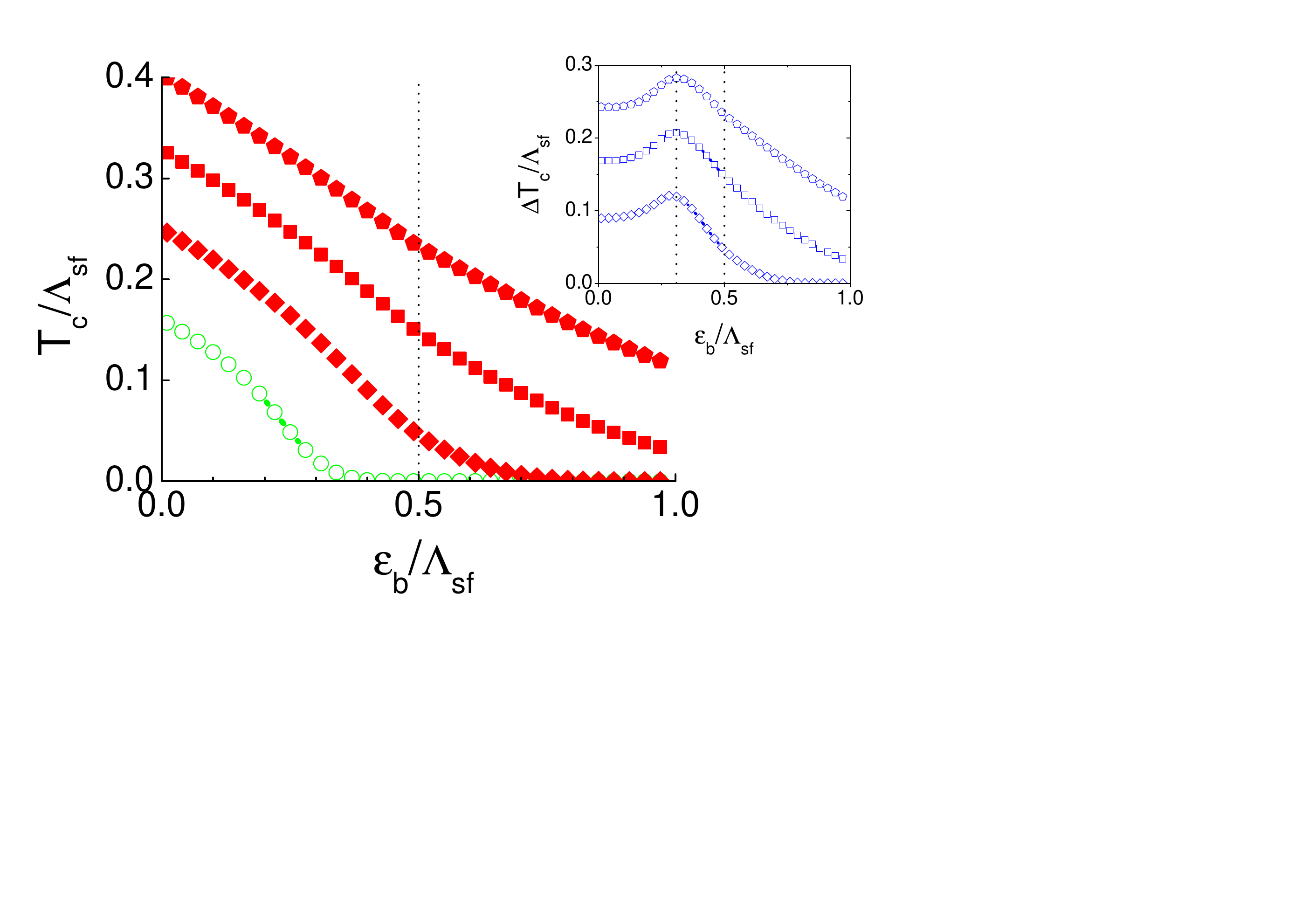}
\vspace{-3cm}
\caption{(Color online) The forward-scattering phonon boost effect. The repulsive spin-fluctuation mediated interaction $\sqrt{N_e N_h} V^{he (eh)}_{sf}=1.5$ and $N_{e(h)} V^{ee (hh)}_{sf}=0.4$ is fixed for all calculations. The forward-scattering phonon interaction is varied as $N_e V^{ee (hh)}_{ph} =0.0, -0.5, -1.0, -1.5$, respectively, in increasing order of $T_c$, and the inter-band phonon interaction $\sqrt{N_h N_e} V^{he (eh)}_{ph}=0.0$ for all cases.
The inset is the plot of the net phonon boost effect $\Delta T_c = T_c -T_c^0$. Vertical lines of $\epsilon_b = 0.32 \Lambda_{sf}$ and $\epsilon_b = \omega_D=0.5\Lambda_{sf}$ are guides for eyes.\label{fig3}}
\end{figure}

\begin{figure}
\noindent
\includegraphics[width=120mm]{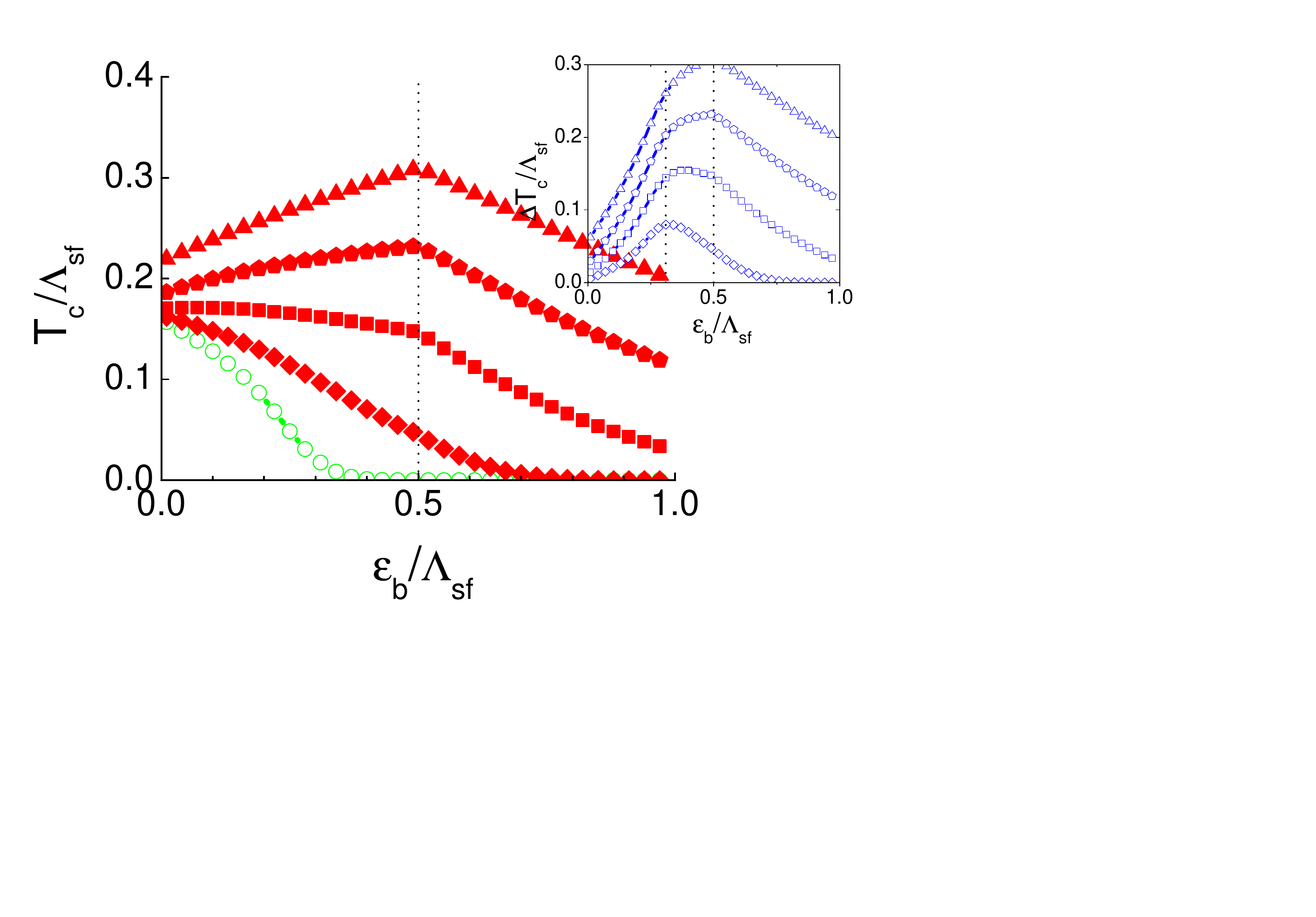}
\vspace{-3cm}
\caption{(Color online) The all-momentum-scattering phonon boost effect. The repulsive spin-fluctuation mediated interaction $\sqrt{N_e N_h} V^{he (eh)}_{sf}=1.5$ and $N_{e(h)} V^{ee (hh)}_{sf}=0.4$ is fixed for all calculations. The all-momentum-scattering phonon interaction, $N_e V^{ee (hh)}_{ph}=\sqrt{N_h N_e} V^{he (eh)}_{ph}=\lambda_{ph}$, is varied as $\lambda_{ph} =0.0, -0.5, -1.0, -1.5$, and $-2.0$, respectively, in increasing order of $T_c$. The inset is the plot of the net phonon boost effect $\Delta T_c = T_c -T_c^0$
\label{fig4}}
\end{figure}

\section{Summary and Conclusions}

We have studied the phonon boost effect on the $s_{he}^{+-}$-pairing state of the incipient two band model. We have considered both the forward-scattering phonon and the all-momentum-scattering local phonon.
It is confirmed that the forward-scattering phonon is efficient to boost $T_c$. In detail, we found that the net phonon boost effect $\Delta T_c= T_c - T_c^0$ has an interesting dependence on $\epsilon_b$ and has a maximum peak near $\epsilon_b^{\ast}$ where $T_c^0$ approaches zero. Our model calculations also demonstrated that the optimal condition of the forward-scattering phonon for increasing $T_c$ is to limit the "forwardness" narrower than the inter-band distance but wide enough to cover the FSs of each of the electron and hole band.

The most important result of our study is that the all-momentum-scattering phonon can be as effective as the forward-scattering phonon to increase $T_c$. This surprising result is, in fact, a natural consequence of the intrinsic property of the incipient band model, where a normal band (crossing the Fermi level) and an incipient band (sunken below the Fermi level) severely break the balance between the gap sizes of $\Delta_h^{+}$ and $\Delta_e^{-}$ on each band. This severe gap size disparity (including the pair susceptibility $\chi_{ph}^{(h(e)}$) turns the all-momentum-scattering phonon into an effective forward-scattering phonon. Since this disparity of the gap size is growing as the incipient band sinks deeper, the phonon-boost effect of the all-momentum-scattering phonon increases as $\epsilon_b$ increases until $\epsilon_b$ reaches to $\epsilon_b^{\ast}$ or $\omega_D$ depending on the phonon coupling strength.

In conclusion, we have shown the theoretical principle how the incipient band turns the local all-momentum-scattering phonon into the effective forward-scattering phonon. Our finding has an important implication that all interface phonons -- not only the 90 $meV$ F-K phonon but also the 60 $meV$ F-K phonon -- from the STO substrate as well as all intrinsic phonons inside the FeSe-layer itself should contribute to increasing $T_c$ of the FeSe/STO monolayer system, if the pairing gap symmetry is the $s_{he}^{+-}$-wave state.
Besides the FeSe/STO monolayer system, other heavily electron-doped iron selenide (HEDIS) compounds such as A$_x$Fe$_{2-y}$Se$_2$ (A=K, Rb, Cs, Tl, etc.) ($T_c \approx 30-40K$)\cite{HEDIS1,HEDIS2,HEDIS3} and (Li$_{1-x}$Fe$_x$OH)FeSe ($T_c \approx 40K$) \cite{OHFeSe}, which develop a deeply sunken ($\epsilon_b \sim 60meV - 90 meV$) incipient band by electron doping, should also have the phonon-boost effect from the intrinsic phonons in the bulk.

{\it Acknowledgements -- } This work was supported by NRF Grant
2016-R1A2B4-008758 funded by the National Research Foundation of
Korea.


\begin{references}
\bibitem{exception}
An unusual phonon such as a backward-scattering phonon, that is stronger for the large-momentum exchange, would suppress $T_c$. However, we do not consider this case.
\end{references}
\end{document}